# Fast Heuristic Algorithm for Multi-scale Hierarchical Community Detection

Eduar Castrillo
Departamento de Ingeniería de Sistemas
Universidad Nacional de Colombia
Bogotá D.C., Colombia
emcastrillov@unal.edu.co

Elizabeth León
Departamento de Ingeniería de Sistemas
Universidad Nacional de Colombia
Bogotá D.C., Colombia
eleonguz@unal.edu.co

Jonatan Gómez
Departamento de Ingeniería de Sistemas
Universidad Nacional de Colombia
Bogotá D.C., Colombia
jgomezpe@unal.edu.co

*Abstract*—Complex networks constitute the backbones of many complex systems such as social networks. Detecting the community structure in a complex network is both a challenging and a computationally expensive task. In this paper, we present the HAMUHI-CODE, a novel fast heuristic algorithm for multi-scale hierarchical community detection inspired on an agglomerative hierarchical clustering technique. We define a new structural similarity of vertices based on the classical cosine similarity by removing some vertices in order to increase the probability of identifying inter-cluster edges. Then we use the proposed structural similarity in a new agglomerative hierarchical algorithm that does not merge only clusters with maximal similarity as in the classical approach, but merges any cluster that does not meet a parameterized community definition with its most similar adjacent cluster. The algorithm computes all the similar clusters at the same time is checking if each cluster meets the parameterized community definition. It is done in linear time complexity in terms of the number of cluster in the iteration. Since a complex network is a sparse graph, our approach HAMUHI-CODE has a super-linear time complexity with respect to the size of the input in the worst-case scenario (if the clusters merge in pairs), making it suitable to be applied on large-scale complex networks. To test the properties and the efficiency of our algorithm we have conducted extensive experiments on real world and synthetic benchmark networks by comparing it to several baseline state-of-the-art algorithms.

## I. INTRODUCTION

Complex networks form the backbones of many natural and man-made systems such as the internet, social networks, protein-protein interaction networks, technological networks and citation networks. These systems have non-trivial topological features and have common properties such as *small-world*, *scale-free* and *community structure* [1]. The study of the complex networks and their properties has become an important research area in disciplines such as mathematics, biology, statistics, computer science, etc. [2].

Detecting the community structure can help us to understand the topological structure and behavior of the network. In [3] a community is defined intuitively as a group of vertices with a high density of connections among members of the group and low density of connections with the rest of the network. No definition of community is universally accepted, and giving a formal or quantitative definition is a challenging task. In fact, the actual definition of community depends on the algorithm, for that reason, several notions of community have been proposed [7].

In the past two decades, various algorithms have been developed for community detection. These algorithms can be categorized according to their functionality as hierarchical clustering, graph partitioning, partitional clustering, spectral methods, and optimization techniques among others. We refer to [7] for a detailed description and a comparative analysis of each category.

Algorithms for community detection with high quality results in small or medium sized networks are not suitable to work with large networks due to their high computational complexity (e.g., Agglomerative Hierarchical Walktrap [25] and Divisive Hierarchical GN [26]). With the arrival of the Big Data era a vast amount of data is generated continuously. For example, many large-scale complex networks with thousands and millions of vertices and edges are generated from different sources such as social networks, peer-to-peer networks, the internet, etc. [15, 16] Algorithms are then required to handle data efficiently, so it is necessary to design and develop efficient methods to handle community detection on big networks.

This paper presents a novel fast heuristic algorithm for multi-scale hierarchical community detection that can be applied efficiently to large-scale complex networks. The algorithm finds the community structure by clustering the vertices inspired on an agglomerative hierarchical algorithm composed of three steps. In the first step, a similarity measure for each edge is computed using a modified version of the structural similarity measure based on vertex neighborhood [8]. In the second step, a heuristic builds meaningful communities as follows: each vertex is set in a separate community, then per iteration, each community $u$ is merged with its most similar adjacent community $v$ when $u$ does not meet a specified community definition. This procedure iterates until all detected communities meet the specified community definition. In the third step, a heuristic merges communities as follows: per iteration, each community $u$ is merged with its most similar adjacent community $v$ when the $u$'s size is less than a specified size. This procedure iterates until all detected communities achieve the minimum size required. Finally, the detected community structure is returned.

The proposed heuristic algorithm can uncover relevant community structure in complex networks, and also presents some interesting properties: *i)* It achieves a time complexity of $O(|E|)$ in the average-case scenario. Such property makes it suitable to handle large-scale complex networks in a reasonable amount of time; *ii)* It can detect communities at any scale due to it does not possess a resolution limit, and it is able to detect hierarchical community structure in a network by tuning an intuitive parameter; *iii)* The algorithm depends on two self-descriptive parameters, the minimum community size $k$ and the community definition $c$. These parameters with two default values are enough to detect a first hierarchy of communities. Furthermore, the algorithm is unsupervised since we do not need to specify the numbers of communities; *iv)* As opposed to the classical Agglomerative Hierarchical Clustering, per iteration, our algorithm can merge not only the clusters with maximal/minimal similarity, but any cluster that does not meet a specified community definition, and also it can achieve the convergence before a single cluster remains.

This paper is divided into 5 sections. Section 2 shows the related work in the literature to our algorithm. Section 3 shows the proposed algorithm. Section 4 presents the experimentation in sythetic and real-world networks by comparing our algorithm to baseline state-of-the-art algorithms. Section 5 draws some conclusions.

## II. RELATED WORK

Many algorithms have been developed to perform community detection by using different approaches (mentioned in Section I). But the approach based on agglomerative and/or hierarchical techniques has shown good results in terms of quality of the detected community structure and the speed of computation. Next, we describe baselines and recent works related to this approach.

Optimization Techniques: The idea behind these methods is that a good community structure must present high values of the modularity measure [20]. Fast Greedy [24] performs greedy modularity optimization with an agglomerative hierarchical approach. Multilevel [6] is a fast modularity optimization algorithm that performs an agglomerative hierarchical approach composed of two phases: network collapse and greedy optimization. These two algorithms find good local optima of the modularity. However, it has been proved that optimizing the modularity yields to the problem of resolution limit [23], making the modularity-based methods unable to detect communities smaller than a certain size that depends on the size of the network. Infomap [21] applies a greedy technique to minimize an objective function called the map equation. The map equation quantifies the information needed to represent a random walker in a network using a two-level nomenclature. Experimentally, Infomap has shown high computational complexity in large-scale complex networks.

SCAN-based Methods: The SCAN algorithm and its variants [8, 9, 10] cluster dense zones of vertices determined by the structural similarity of vertices. They are fast but strongly depend on a minimum similarity parameter $\varepsilon$ that is difficult to estimate. To overcome the problem of estimating $\varepsilon$, the parameter-less algorithms SHRINK-H and SHRINK-G [4] were proposed. SHRINK-H performs agglomerative hierarchical clustering by merging dense pairs into micro-communities that conforms the hierarchy of communities, and the final clustering is determined by the partition that maximize the modularity. SHRINK-H presents a worst-case time complexity making it unable to handle efficiently large datasets. SHRINK-G is faster but it sacrifices the capacity of finding hierarchical community structure. AHSCAN [11] performs agglomerative hierarchical clustering by iteratively merging pair of vertices in order of decreasing structural similarity of vertices, until a single cluster remains. AHSCAN selects the partition that maximizes the modularity, so its time complexity scales by $O(|E||V|)$.

Hierarchical Methods: Dense Sub Graph Extraction (DSE) [12] uses a combination of the matrix blocking technique and hierarchical clustering for extracting dense subgraphs defined by the structural similarity of vertices. DSE is fast and presents high quality results, but it requires as parameter the minimum density of communities, which is not easy to estimate, because the density of communities may vary in the same network. Recently, a fast algorithm was proposed in [13]. It uses a two-step method to build the community structure based on label propagation. In the first step, the structural similarity of vertices is computed for each pair of adjacent vertices. Additionally, it applies the label propagation algorithm to detect meta-communities conformed of most similar vertices. In the second step, a multilevel label propagation technique is applied to build communities that meet a modified version of the Weak community definition [3]. The second step is based on two sub-steps: network collapse and label propagation. The algorithm introduces the cohesion parameter $\lambda$ (a real value within the range [0, 1]) to control the resolution of the resulting communities.

Linear Time Algorithms: Label Propagation proposed by Raghavan et al. [22] simulates the diffusion of information (labels) through the network. At the beginning, each vertex is labeled with a unique value, then iteratively each vertex takes the most frequent label in its neighborhood and the process continues until convergence. The results provided by Label Propagation are sometimes unpredictable and the whole network can be detected as a single partition. Other recently proposed algorithm is Attractor [18] that investigates local distance dynamics among connected vertices. Attractor computes the distance on edges based on the Jaccard similarity and applies 3 interaction patterns until the distances converge. The resulting communities are the connected components generated through the removal of the edges with final distance of one. Attractor introduces a cohesion parameter $\lambda$ (a real value within the range [0, 1]) to control the resolution of the detected communities.

## III. COMMUNITY DETECTION ALGORITHM

The community detection problem has been considered as the discovering of dense sub graphs defined by the structural similarity of vertices. This similarity has proved to be very effective in modeling such dense sub graphs [4, 8, 11, 12, 13]. We propose a novel fast heuristic algorithm inspired on an agglomerative hierarchical technique by using this structural similarity measure. This algorithm uncovers the underlying community structure by merging vertices and communities with its most similar adjacent pair.

*A. Definitions*

We will refer to a complex network modeled by a graph $G = (V, E)$ where $V$ is the set of vertices, $E$ is the set of edges connecting pair of vertices $(v, u) \in E$.

*1) Community Detection*: Given a graph $G = (V, E)$, community detection consist in partitioning the set of vertices $V$ into $k$ subsets $P = \{C_1, C_2, ..., C_k\}$ such that $\{C_1 \cup C_2 \cup ... \cup C_k\} = V$ and $\{C_1 \cap C_2 \cap ... \cap C_k\} = \emptyset$.

*2) Vertex Structure*: Let $v \in V$, the structure of vertex $v$ [8] denoted by $\Gamma(v)$

$$\Gamma(v) = \{u \in V \mid (v, u) \in E\} \cup \{v\} \quad (1)$$

*3) Structural Similarity (A.K.A. Cosine Similarity)*: In the structural similarity definition in [8], it is not possible to obtain a minimum similarity $\sigma(v, u) = 0$ when $v$ and $u$ do not share any common neighbors, due to $|\Gamma(v) \cap \Gamma(u)| > 0$. The value $\sigma(v, u)$ can in some cases be a large value depending on the sizes $|\Gamma(v)|$ and $|\Gamma(u)|$. For that reason, we propose a modified definition presented in (2). Let $(v, u) \in E$, the structural similarity denoted by $\sigma(v, u)$

$$\sigma(v, u) = \frac{|\Gamma(v) \cap \Gamma(u) - \{v, u\}|}{\sqrt{|\Gamma(v) - \{u, v\}| \cdot |\Gamma(u) - \{v, u\}|}} \quad (2)$$

This definition returns a minimal structural similarity $\sigma(v, u) = 0$, in the case that $v$ and $u$ do not share any common neighbors, by removing the vertices $\{u, v\}$ from $\Gamma(v)$ and $\Gamma(u)$. This modification increases the probability of identifying inter-community edges when applying the proposed heuristics.

*1) Weak Community*: Let $k_v$ be the degree of a vertex $v$, i.e., the number of adjacent vertices to $v$. If $C \subset V$ is a sub graph to which vertex $v$ belongs, then we can compute $k_v = k_v^{in}(C) + k_v^{out}(C)$, where $k_v^{in}(C)$ is the number of edges connecting $v$ to others vertices in $C$ and $k_v^{out}(C)$ is the number of edges connecting $v$ to the rest of the network. The sub graph $C$ is a community in the Weak sense [3] if

$$\sum_{v \in C} k_v^{in}(C) \geq \sum_{v \in C} k_v^{out}(C)$$

*1) Weakest Community*: Let $P = \{C_1, C_2, ..., C_k\}$ be a partition of $V$ into $k$ subsets. The sub graph $C_i$ is a community in the Weakest sense [5] if

$$\sum_{v \in C_i} k_v^{in} \geq \max_{C_j \in P, C_i \neq C_j} |\{(v, u) \mid (v, u) \in E \wedge u \in C_j\}|$$

We have selected the Weak and Weakest community definitions for the following reasons: *i)* if a random network (Erdős-Rényi model [1] for example) is divided into two disjoint groups of vertices, there is a low probability that the two groups fulfills the community definition [3]; *ii)* the Weak and Weakest definitions are local and independent of the size of the network. An appropriate local heuristic, that uses the selected community definitions as stop criteria, can build communities without resolution limit; *iii)* the Weak and Weakest definitions are compatible with other community definitions. For example, a community defined as Strong [3] can be also defined as Weak, while the opposite is not always true. The same property applies to other Strong and Weak definitions [5].

The advantage of the selected structural similarity and community definitions is that they can be further extended to weighted graphs [3, 4, 5].

*B. The HAMUHI-CODE*

The **H**euristic **A**lgorithm for **MU**lti-scale **HI**erarchical **CO**mmunity **DE**tection (HAMUHI-CODE) proposed is based on the following hypotheses:

Hypothesis 1: If a vertex $u$ presents a maximal similarity measure with an adjacent vertex $v$, then $u$ is more likely to be in the same community as $v$.

Hypothesis 2: If a community $C_1$ presents a maximal similarity measure with an adjacent community $C_2$ through an edge directly connecting $C_1$ and $C_2$, then $C_1$ and $C_2$ are more likely to be part of the same community in the next hierarchical level. Whether to merge $C_1$ to $C_2$ depends exclusively of the current state of $C_1$.

Based on the two hypotheses, the proposed algorithm finds the community structure inspired on an Agglomerative Hierarchical Clustering (AHC) composed of three steps.

Step 1. The structural similarity for each edge is computed once by using (2).

Step 2. The first hierarchical level is computed as follows: each vertex is set in a separate community, then by hypotheses 1 and 2, per iteration, each community $u$ is merged with its most similar adjacent community $v$ when $u$ does not meet a parameterized community definition (instead of merging only the communities with maximal/minimal similarity, as in the classical AHC). This procedure iterates until all detected communities meet the parameterized community definition (instead of iterating until a single community remains, as in the classical AHC).

Step 3. Similarly to Step 2, a next hierarchical level is computed as follows: by hypothesis 2, each community $u$ is merged with its most similar adjacent community $v$ when the $u$'s size is less than a parameterized size. This procedure iterates until all detected communities achieve the minimum size required.

We assume without loss of generality that vertices in $G$ are numbered from zero to $|V| - 1$; otherwise, the vertice's identifiers must be normalized before applying the algorithm. Additionally, we represent the community structure using a disjoint-set data structure in order to perform the following operations optimally: vertex's community querying and adjacent communities merging. We present the detailed pseudo code of a fast and basic implementation of the proposed heuristics in Algorithms 1, 2 and 3.

HAMUHI-CODE (See Algorithm 1): It is the main of the algorithm. At line 1, the similarity for each edge is computed. At line 2 all vertices are assigned into separate communities. At line 3 the community structure with the required community definition $c$ is computed, and in line 4 the hierarchical level is selected.

**Algorithm 1. HAMUHI-CODE**

**Input:** $G = (V, E), k, c$
**Output:** $C$
1: **for each** $(v, u) \in E$ compute $\sigma(v, u)$ using (2)
2: **let** $C$ = disjoint-set of size $|V|$. // $C_i$ = community of vertex $i$
3: COMMUNITYDETECTION($G, C, \sigma, c$)
4: HIERARCHICALLEVEL($G, C, \sigma, k$)

COMMUNITYDETECTION (See Algorithm 2): Builds a community structure where each community meets the parameterized community definition. The array position $W_c$ stores the maximal structural similarity found so far between the community $c$ and an adjacent community stored in the array position $R_c$. If the target community definition is WEAK, the array position $B_c$ stores the difference between the internal and external degree of the community $c$. In the case $B_c < 0$, then the community $c$ does not meet the Weak definition. From line 17 to 19, each community $c$ that does not meet the Weak definition is merged with its most similar adjacent community. If the target community definition is WEAKEST, the array position $B_c$ stores the internal degree of the community $c$ and the array position $D_c$ stores the maximal external degree of the community $c$ towards an adjacent community. At line 21 a hash table $H$ is defined. The keys $H_{i,j}$ and $H_{j,i}$ store the external degree between adjacent communities $i$ and $j$. If $B_c < D_c$, then the community $c$ does not meet the Weakest definition. From line 39 to 41, each community $c$ that does not meet the Weakest definition is merged with its most similar adjacent community.

HIERARCHICALLEVEL (See Algorithm 3): Builds communities having a size no less than $k$. The arrays $W$ and $R$ have the same meaning of that in Algorithm 2. Additionally, the array position $S_c$ stores the current size of the community $c$. From line 12 to 14, each community $c$ whose current size $S_c$ is less than $k$ is merged with its most similar adjacent community.

*C. Complexity Analysis*

In Algorithm 1, computing the structural similarity for each edge in the graph (line 1) takes worst-case time complexity of $O(|E| \cdot \alpha(G))$ where $\alpha(G)$ is the arboricity of $G$ [14]. Algorithms 2 and 3 present worst-case time complexity of $O(|E| \cdot t)$ where $t$ is the number of iterations needed until convergence. At each iteration in these algorithms, the number of communities waiting for convergence is decreased approximately to the half, then $t$ is bounded by $t = O(\log|V|)$. Thus, the worst-case time complexity of HAMUHI is $O(|E| \cdot \alpha(G) + |E| \cdot \log|V|)$. However, real complex networks are usually sparse, i.e., $|E| = O(|V|)$, and exhibit community structure. In these networks, for the arboricity of $G$ we have $\alpha(G) << 0.5 \cdot (2 \cdot |E| + |V|)^{0.5}$, and also a small number of iterations are required to achieve the convergence in Algorithms 2 and 3, obtaining an average time complexity of $O(|E|)$. The space complexity is $O(|E| + |V|)$.

**Algorithm 2. COMMUNITYDETECTION**

**Input:** $G = (V, E), C, \sigma, comdef$
1: **let** $W, R, B, D$ be four arrays of size $|V|$ and **let** $flag$ = True
2: **if** *comdef* **is** WEAK **then**
3:   **while** *flag* **is True do**
4:     *flag* = False
5:     **for each** $c$ **in** $0...|V| - 1$ **do**
6:       $W_c = -1, R_c = -1, B_c = 0$
7:     **for each** $(v, u) \in E$ **do**
8:       **if** $C_v \neq C_u$ **then**
9:         $B_{Cu} = B_{Cu} - 1$
10:         $B_{Cv} = B_{Cv} - 1$
11:         **if** $\sigma(v, u) > W_{Cu}$ **then**
12:           $W_{Cu} = \sigma(v, u), R_{Cu} = C_v$
13:         **if** $\sigma(v, u) > W_{Cv}$ **then**
14:           $W_{Cv} = \sigma(v, u), R_{Cv} = C_u$
15:       **else then**
16:         $B_{Cu} = B_{Cu} + 2$
17:     **for each** $c$ **in** $0...|V| - 1$ **do**
18:       **if** $B_c < 0$ **and** $R_c \neq -1$ **then**
19:         $C.merge(c, R_c)$ and $flag$ = True
20: **else if** *comdef* **is** WEAKEST **then**
21:   **let** $H$ be a hash table of size $|E|$
22:   **while** *flag* **is True do**
23:     *flag* = False
24:     **for each** $c$ **in** $0...|V| - 1$ **do**
25:       $W_c = -1, R_c = -1, B_c = 0, D_c = 0$
26:     **for each** $(v, u) \in E$ **do**
27:       **if** $C_v \neq C_u$ **then**
28:         $H_{Cu, Cv} = H_{Cu, Cv} + 1$
29:         **if** $H_{Cu, Cv} \geq D_{Cu}$ **then**
30:           $D_{Cu} = H_{Cu, Cv}$
31:         **if** $H_{Cu, Cv} \geq D_{Cv}$ **then**
32:           $D_{Cv} = H_{Cu, Cv}$
33:         **if** $\sigma(v, u) > W_{Cu}$ **then**
34:           $W_{Cu} = \sigma(v, u), R_{Cu} = C_v$
35:         **if** $\sigma(v, u) > W_{Cv}$ **then**
36:           $W_{Cv} = \sigma(v, u), R_{Cv} = C_u$
37:       **else then**
38:         $B_{Cu} = B_{Cu} + 2$
39:     **for each** $c$ **in** $0...|V| - 1$ **do**
40:       **if** $B_c < D_c$ **and** $R_c \neq -1$ **then**
41:         $C.merge(c, R_c)$ and $flag$ = True

**Algorithm 3. HIERARCHICALLEVEL**

**Input:** $G = (V, E), C, \sigma, k$
1: **let** $W, R, S$ be three arrays of size $|V|$ and **let** $flag$ = True
2: **while** *flag* **is True do**
3:   *flag* = False
4:   **for each** $c$ **in** $0...|V| - 1$ **do**
5:     $W_c = -1, R_c = -1, S_c = C.size(c)$
6:   **for each** $(v, u) \in E$ **do**
7:     **if** $C_v \neq C_u$ **then**
8:       **if** $\sigma(v, u) > W_{Cu}$ **then**
9:         $W_{Cu} = \sigma(v, u), R_{Cu} = C_v$
10:       **if** $\sigma(v, u) > W_{Cv}$ **then**
11:         $W_{Cv} = \sigma(v, u), R_{Cv} = C_u$
12:   **for each** $c$ **in** $0...|V| - 1$ **do**
13:     **if** $S_c < k$ **and** $R_c \neq -1$ **then**
14:       $C.merge(c, R_c)$ and $flag$ = True

Fig. 1 shows an illustration of the execution of HAMUHI($k$=2, $c$=WEAK) on a sample network. As we can notice, a partition composed of Weak communities is computed by applying a single iteration of the algorithm. In fact, the computed partition corresponds to the final community structure.

## IV. EXPERIMENTS

We have conducted experiments with HAMUHI on synthetic and real-world networks. In doing so, we tested the properties, efficacy and efficiency of our proposal. All the experiments were running in a machine with 32 GB RAM and a 3.4 GHz CPU. A single-threaded version of HAMUHI was implemented in C++11.

### A. State-of-the-art Algorithms

We have selected a set of state-of-the-art algorithms to be compared with HAMUHI. The selected algorithms are Fast Greedy (FG), Multilevel (ML), Infomap (IM) and Label Propagation (LP). We used the implementations (in C language) of these algorithms provided in [17]. The selected algorithms are baseline and representative in community detection because of their high quality results and mainly because of their low time complexity.

### B. Evaluation Measures

To test the effectiveness of HAMUHI in the case of networks with ground truth communities, we have selected the *Normalized Mutual Information* (NMI) used in [13]. NMI compares two partitions generated from the same dataset by assigning a score within the range [0, 1], where 0 indicates that the two partitions are independent from each other and 1 if they are equal. If ground truth information is not provided, we measure the performance of HAMUHI with the *Modularity Q* [20].

### C. Sensitivity to the Similarity Measure

Fig. 2 shows the results of executing HAMUHI with the original and the modified version of the structural similarity on the Bottlenose Dolphins dataset [16]. This network is composed of 62 vertices and 159 edges and in the ground truth the vertices are divided into two groups.

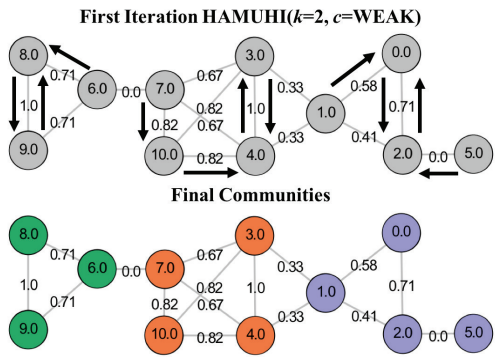

Fig. 1. Result of HAMUHI on a sample network. The edge's weight corresponds to the computed structural similarity. (Top) The arrows indicate the merge operations performed in a single iteration of the Algorithm 2. (Bottom) Colors represent the communities discovered after the first iteration.

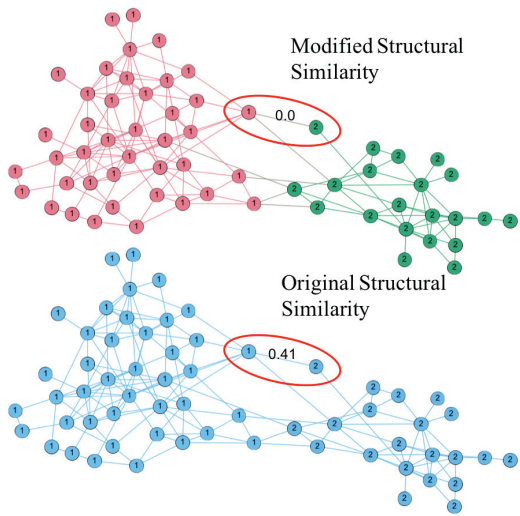

Fig. 2. Results of HAMUHI on dolphins dataset [16]. Colors represent communities discovered by HAMUHI and numbers represent ground-truth communities. With the original similarity, the algorithm fails to detect the ground truth, due to the problematic edge highlighted in the ellipse.

As we can see, the algorithm fails to detect the ground-truth by employing the original definition of the structural similarity, due to the problematic edge highlighted in the ellipse. In that edge, its endpoints do not share any common neighbors, but they merge together because the computed similarity $\sigma = 0.41$ results maximal for both of them. On the other hand, by employing the modified structural similarity, the problematic edge is computed to $\sigma = 0$, then it is treated as a potential inter-community edge. The modified version also outperforms the strategy of calculating the original similarity for all the edges and setting $\sigma = 0$ for those edges whether its endpoints do not share any common neighbors.

### D. LFR Benchmark

The *LFR* benchmark [19] generates unweighted and undirected graphs with ground truth. Also, it produces networks with vertex degree and community size that follow power law distributions, making it more appropriate than the GN benchmark [24] to model complex networks. By varying the mixing parameter $\mu$, LFR can generate networks with community structure more or less difficult to identify.

We ran an experiment using the parameters listed in Table I. For this experiment we have executed the algorithms over 50 runs of the LFR benchmark for each combination of parameters, and the mean and standard deviation of the NMI were calculated. As we can notice in Fig. 3, the critical point on the performance, for the majority of the algorithms arrives when the mixing parameter $\mu > 0.5$. LP (Fig. 3 (e)) presents good performance until $\mu \sim 0.4$ where it becomes particularly erratic, presenting high standard deviation. In contrast, FG (Fig. 3 (f)) offers stable results but they are not so good because of their low values of NMI. On the other hand, the ML's (Fig. 3 (d)) performance is affected notably when the size of the network increase due to its resolution limit. IM (Fig. 3 (c)) is the best performant algorithm before $\mu = 0.6$, offering almost perfect and stable results, but after $\mu > 0.6$ it shows an abrupt decay in its performance. HAMUHI($k$=2, $c$=WEAK)

(Fig. 3 (a)) performs near to the optimal with stable results while $\mu < 0.5$, but it presents an abrupt change of behavior after the critical point, because for values of $\mu > 0.5$ no community in the generated ground truth meets the Weak definition (it makes the ground truth undetectable). However, it outperforms the other algorithms before the critical point, with the exception of IM. In contrast, HAMUHI($k$=2, $c$=WEAKEST) (Fig. 3 (b)) exhibits a minimum loss of accuracy before $\mu = 0.5$ with respect to its other version, but it presents a smooth transition after the critical point outperforming the other algorithms in the majority of scenarios for values $\mu \geq 0.6$.

### E. Sensitivity to the Parameterization

We have tested HAMUHI($k$, $c$) on a set of real-world networks listed in Table II, by varying the parameter $k$ in the interval [2, 500] and the parameter $c$ in the values {WEAK, WEAKEST}. Fig 4. shows the modularity $Q$ computed for each combination of parameters-networks. As we can notice, by increasing the parameter $k$, the tendency is to obtain partitions with higher values of modularity because the small communities merge into larger ones. For the Ego-Facebook network, the modularity decrease on intervals for values of $k > 200$, possibly because the network becomes small with respect to the value of $k$. Moreover, all the resulting partitions present good modularity rates, since their values are above 0.6. If higher values of modularity are desired as result, then sampling values of $k$ can be a good search strategy. However, with a default value of $k$=2 it is enough to detect an initial relevant partition. An equal tendency was obtained for $c$=WEAKEST.

Fig. 5 shows the number of detected communities for each combination of parameters-networks. In each case, the number of communities decreases following a power law distribution. This result is related to the community size distribution obtained in these complex networks, that also follows a power law distribution (See Section IV-F).

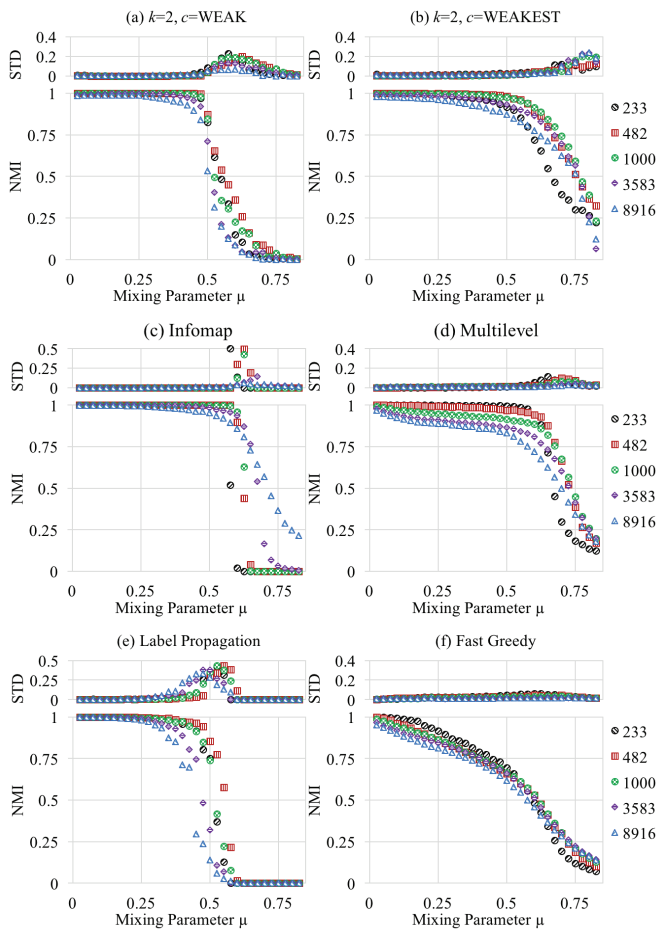

Fig. 3. (Lower row) The mean value of the Normalized Mutual Information, NMI (higher values are better) as a function of the mixing parameter $\mu$. (Upper row) The standard deviation, STD (lower values are better) of the NMI as a function of $\mu$. The parameters of the experiment are detailed in Table I.

TABLE I. PARAMETERS FOR THE LFR BENCHMARK

| Parameter | Value |
|---|---|
| Number of vertices N | {233, 482, 1000, 3583, 8916} |
| Maximum degree | 0.1N |
| Maximum community size | 0.1N |
| Average degree | 20 |
| Degree distribution exponent | -2 |
| Community size distribution exponent | -2.5 |
| Mixing parameter | [0.025, 0.825] with 0.025 step |

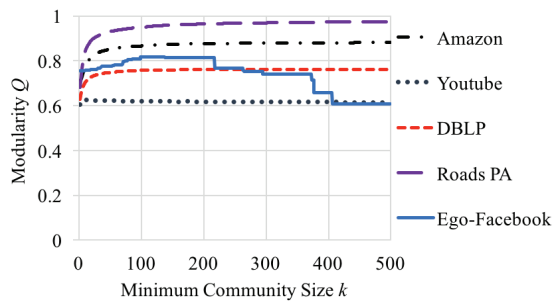

Fig. 4. Modularity rate $Q$ (Higher values are better) in function of HAMUHI($k$, $c$=WEAK) for different values of $k$.

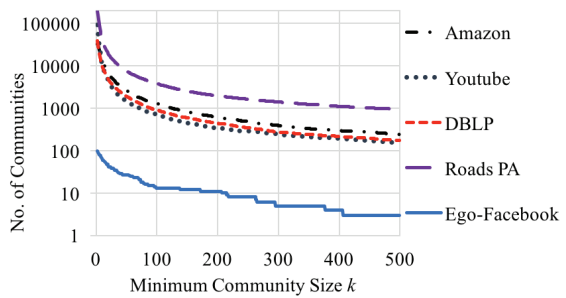

Fig. 5. Number of detected communities (in log-scale) in function of HAMUHI($k$, $c$=WEAK) for different values of $k$.

TABLE II. LARGE REAL-WORLD NETWORKS EXTRACTED FROM [17]

| Network | $|V|$ | $|E|$ |
|---|---|---|
| Amazon | 334,863 | 925,872 |
| YouTube | 1,134,890 | 2,987,624 |
| DBLP | 317,080 | 1,049,866 |
| Roads PA | 1,088,092 | 1,541,898 |
| Ego-Facebook | 4,039 | 88,234 |

## F. Multi-scale Community Detection

Fig. 6 shows the community size power-law distribution obtained by HAMUHI($k$=2, $c$=WEAKEST) in both, the YouTube and Amazon networks [15]. This result indicates that HAMUHI can detect communities at any scale, from very small communities to very larges ones, independent of the size of the network. This capacity lacks on algorithms that suffer of resolution limit (e.g., Multilevel).

## G. Hierarchical Community Detection

We have tested our algorithm in a hierarchical complex network that follows the Ravasz-Barabási model [27]. The network is composed of two hierarchical levels, with 25 communities (5 vertices per community) in the first level, and 5 communities in the second level. Fig. 7 shows the two levels identified by HAMUHI($k$=2, $c$=WEAKEST) and HAMUHI($k$=6, $c$=WEAKEST). The ground truth communities were correctly identified in both levels. A possible strategy to detect the next hierarchical level in a network is by setting the parameter $k$ to $k_{level+1}$ = MinCommunitySize$_{level}$ + 1.

## H. Random Networks

We tested our algorithm on random networks that follow the Erdős-Rényi model and Scale-Free Barabási-Albert model [1]. In the Erdős-Rényi model, each pair of vertices in the graph have the same probability $p$ of being connected by an edge, resulting in a binomial distribution in the vertex degree. In the Barabási-Albert model, a new vertex is joined to other $m$ existing vertices in the graph via preferential attachment, resulting in a power-law distribution in the vertex degree. The networks generated using these two models are expected to have no community structure.

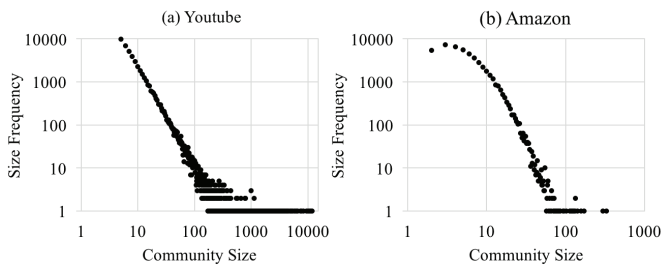

Fig. 6. Community size distribution obtained with HAMUHI($k$=2, $c$=WEAKEST) in the YouTube and Amazon networks [15]. The results fit to power-law distributions with exponents -1.30 and -2.93 respectively.

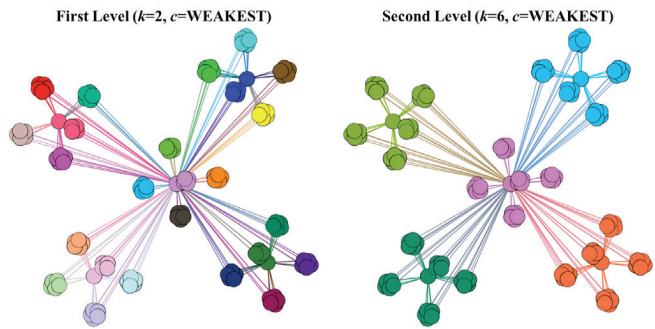

Fig. 7. Communities detected by HAMUHI in a Ravasz-Barabási complex network with two hierarchical levels. The two levels were correctly identified.

For this experiment, the size of the networks in both models was fixed to 1000 vertices. Fig. 8 shows the results of executing the algorithms on random networks. As expected, HAMUHI, LP and IM detect only one big community in all cases.

## I. Resolution Limit

We have built two ring networks in order to check the resolution limit of the algorithms. The ring networks are composed of $N$ identical cliques (3-cliques or 4-cliques) connected by a single edge [23].

Fig. 9 shows the results obtained for each algorithm on the ring networks. HAMUHI is able to identify all cliques as separate communities in both cases. LP is close to detect the majority of communities. IM has problems on identifying the communities in the 3-Clique ring network, but it obtains a perfect score in the 4-Clique ring network. FG and ML have the worst results and it is due to their resolution limit.

## J. Modularity and Running Time on Large-Scale Networks

We have selected some large-scale real-world networks to test HAMUHI. Table III shows the modularity score and the running time obtained after having executed HAMUHI, ML, LP and IM on each network. As we can notice, HAMUHI obtains the best running time on all datasets, proving experimentally its low time complexity. In fact, HAMUHI obtains a low running time in the Facebook-KONECT and Facebook-UCI-UNI networks because the Algorithm 2 requires only 6 and 8 iterations respectively to achieve the convergence. In the case of the Livejournal network, the Algorithm 2 requires 28 iterations to achieve the convergence even though the Livejournal network is smaller (in term of number of vertices and edges) than the two Facebook networks.

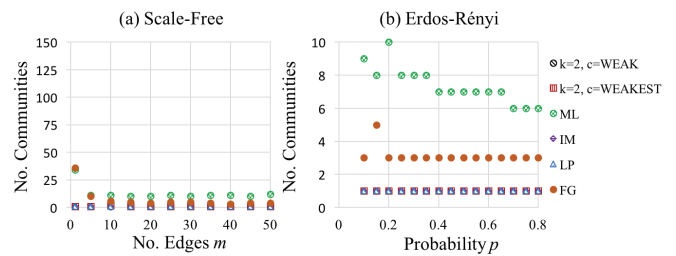

Fig. 8. Number of detected communities in random networks (lower values are better). These networks follow the models Scale-free Barabási-Albert and the Erdős-Rényi.

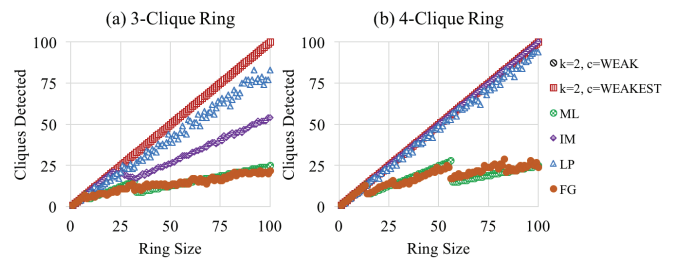

Fig. 9. Test of resolution limit. Number of detected communities (higher values are better) on the ring networks composed of identical cliques connected by a single edge.

TABLE III. RESULTS OF EXECUTING HAMUHI($k$=2, $c$=WEAK) ON LARGE-SCALE SOCIAL NETWORKS

| Network | $|V|$ | $|E|$ | Modularity | | | | Time (seconds) | | | |
|---|---|---|---|---|---|---|---|---|---|---|
| | | | *HAMUHI* | *ML* | *LP* | *IM* | *HAMUHI* | *ML* | *LP* | *IM* |
| Facebook-UCI-UNI [18] | 58,790,782 | 92,208,195 | 0.69 | **0.90** | 0.63 | -- | **75.73** | 629.21 | 32,839.3 | -- |
| Facebook-KONECT [18] | 59,216,211 | 92,522,012 | 0.69 | **0.90** | 0.63 | -- | **71.40** | 468.13 | 39,117.5 | -- |
| Livejournal [18] | 4,033,137 | 27,933,062 | 0.56 | **0.74** | 0.48 | 0.02 | **54.72** | 157.9 | 765.53 | 1766.0 |
| Youtube [17] | 1,134,890 | 2,987,624 | 0.60 | **0.71** | 0.66 | 0.69 | **4.63** | 8.95 | 112.60 | 92.0 |
| Amazon [17] | 334,863 | 925,872 | 0.711 | **0.92** | 0.78 | 0.79 | **0.27** | 5.57 | 25.13 | 35.0 |

a. The results not obtained in a time gap of 40,000 seconds are marked as –

Additionally, while HAMUHI, LP and IM achieve similar modularity values, ML achieves the higher values in all networks, but in practice, the modularity in complex networks lies in the interval [0.3, 0.7] [20]. Higher values are rare and are biased towards resolution limit.

## V. CONCLUSIONS

In this paper we present HAMUHI, a novel fast heuristic algorithm for multi-scale hierarchical community detection in complex networks inspired on an agglomerative hierarchical clustering technique, that merges vertices and communities with high structural similarity. Through extensive experiments we show that our proposal can detect high quality community structure in networks, and compared to several state-of-the-art algorithms, HAMUHI obtains superior results in several scenarios, so it can be considered a good candidate to perform community detection, specially on large-scale complex networks since its time complexity scales by $O(|E|)$ in the average case. Furthermore, it presents an intuitive parameterization, requiring the minimum community size $k$ and the community definition $c$ as parameters. The experimental results have shown that using the default values ($k$=2, $c$=WEAKEST) is enough to detect a first relevant hierarchy of communities in a network. As future work, we plan to extend the algorithm to detect online communities in dynamic networks.

REFERENCES

[1] Newman, M. E. (2003). The structure and function of complex networks. SIAM review, 45(2), 167-256.

[2] Erciyes, K. (2014). Complex networks: an algorithmic perspective. CRC Press.

[3] Radicchi, F., Castellano, C., Cecconi, F., Loreto, V., & Parisi, D. (2004). Defining and identifying communities in networks. Proceedings of the National Academy of Sciences of the United States of America, 101(9), 2658-2663.

[4] Huang, J., Sun, H., Han, J., Deng, H., Sun, Y., & Liu, Y. (2010, October). SHRINK: a structural clustering algorithm for detecting hierarchical communities in networks. In Proceedings of the 19th ACM international conference on Information and knowledge management (pp. 219-228). ACM.

[5] Hu, Y., Chen, H., Zhang, P., Li, M., Di, Z., & Fan, Y. (2008). Comparative definition of community and corresponding identifying algorithm. Physical Review E, 78(2), 026121.

[6] Blondel, V. D., Guillaume, J. L., Lambiotte, R., & Lefebvre, E. (2008). Fast unfolding of communities in large networks. Journal of statistical mechanics: theory and experiment, 2008(10), P10008.

[7] Fortunato, S., & Hric, D. (2016). Community detection in networks: A user guide. Physics Reports, 659, 1-44.

[8] Xu, X., Yuruk, N., Feng, Z., & Schweiger, T. A. (2007, August). SCAN: a structural clustering algorithm for networks. In Proceedings of the 13th ACM SIGKDD international conference on Knowledge discovery and data mining (pp. 824-833). ACM.

[9] Shiokawa, H., Fujiwara, Y., & Onizuka, M. (2015). SCAN++: efficient algorithm for finding clusters, hubs and outliers on large-scale graphs. Proceedings of the VLDB Endowment, 8(11), 1178-1189.

[10] Chang, L., Li, W., Lin, X., Qin, L., & Zhang, W. (2016, May). pSCAN: Fast and exact structural graph clustering. In Data Engineering (ICDE), 2016 IEEE 32nd International Conference on (pp. 253-264). IEEE.

[11] Yuruk, N., Mete, M., Xu, X., & Schweiger, T. A. (2009, July). AHSCAN: Agglomerative hierarchical structural clustering algorithm for networks. In Social Network Analysis and Mining, 2009. ASONAM'09. International Conference on Advances in (pp. 72-77). IEEE.

[12] Chen, J., & Saad, Y. (2012). Dense subgraph extraction with application to community detection. IEEE Transactions on Knowledge and Data Engineering, 24(7), 1216-1230.

[13] Han, J., Li, W., & Deng, W. (2016). Multi-scale community detection in massive networks. Scientific Reports, 6.

[14] Chiba, N., & Nishizeki, T. (1985). Arboricity and subgraph listing algorithms. SIAM Journal on Computing, 14(1), 210-223.

[15] Leskovec, J. & Krevl, A. SNAP Datasets: Stanford large network dataset collection. http://snap.stanford.edu/data (2014).

[16] Rossi, R., & Ahmed, N. (2015, January). The Network Data Repository with Interactive Graph Analytics and Visualization. In AAAI (Vol. 15, pp. 4292-4293).

[17] Csardi, G., & Nepusz, T. (2006). The igraph software package for complex network research. InterJournal, Complex Systems, 1695(5), 1-9.

[18] Shao, J., Han, Z., & Yang, Q. (2014). Community Detection via Local Dynamic Interaction. arXiv preprint arXiv:1409.7978.

[19] Lancichinetti, A., Fortunato, S., & Radicchi, F. (2008). Benchmark graphs for testing community detection algorithms. Physical review E, 78(4), 046110.

[20] Newman, M. E., & Girvan, M. (2004). Finding and evaluating community structure in networks. Physical review E, 69(2), 026113.

[21] Rosvall, M., & Bergstrom, C. T. (2008). Maps of random walks on complex networks reveal community structure. Proceedings of the National Academy of Sciences, 105(4), 1118-1123.

[22] Raghavan, U. N., Albert, R., & Kumara, S. (2007). Near linear time algorithm to detect community structures in large-scale networks. Physical review E, 76(3), 036106.

[23] Fortunato, S., & Barthelemy, M. (2007). Resolution limit in community detection. Proceedings of the National Academy of Sciences, 104(1), 36-41.

[24] Clauset, A., Newman, M. E., & Moore, C. (2004). Finding community structure in very large networks. Physical review E, 70(6), 066111.

[25] Pons, P., & Latapy, M. (2005, October). Computing communities in large networks using random walks. In International Symposium on Computer and Information Sciences (pp. 284-293). Springer Berlin Heidelberg.

[26] Girvan, M., & Newman, M. E. (2002). Community structure in social and biological networks. Proceedings of the national academy of sciences, 99(12), 7821-7826.

[27] Ravasz, E., & Barabási, A. L. (2003). Hierarchical organization in complex networks. Physical Review E, 67(2), 026112.